\documentclass{PoS}

\usepackage{bm} 
\usepackage{graphicx}

\title{
Pion cloud and sea quark flavor asymmetry 
in the impact parameter representation}

\ShortTitle{Pion cloud and sea quark flavor asymmetry}

\author{M.~Strikman\\
        Department of Physics, Pennsylvania State University,
        University Park, PA 16802, USA\\
        E-mail: \email{strikman@phys.psu.edu}}

\author{\speaker{C.~Weiss}\\
        Theory Center, Jefferson Lab, Newport News, VA 23606, USA\\
        E-mail: \email{weiss@jlab.org}}

\abstract{We study large--distance contributions to the 
nucleon parton densities in the transverse coordinate (impact parameter) 
representation based on generalized parton distributions (GPDs). 
Chiral dynamics generates a distinct component of the partonic structure, 
located at momentum fractions $x \lesssim M_\pi / M_N$ and transverse 
distances $b \sim 1/M_\pi$. We analyze the phenomenological ``pion cloud'' 
model of the flavor asymmetry $\overline{d}(x) - \overline{u}(x)$ 
and quantify what fraction of the calculated asymmetry
results from the universal large--distance region. Our findings indicate 
that a two--component picture of the nucleon's partonic structure,
with a ``core'' antiquark distribution at 
$b < b_{\rm core} \approx 0.55 \, \textrm{fm}$ which vanishes at 
$x \rightarrow 0$ and the universal large--distance pion cloud, 
could naturally account for the $x$--dependence of the measured asymmetry.}

\FullConference{LIGHT CONE 2008 Relativistic Nuclear and Particle Physics\\
		 July 7-11  2008\\
		 Mulhouse, France}

\begin{document}

\section{Introduction}
Experiments in deep--inelastic lepton--nucleon scattering 
\cite{Amaudruz:1991at} and Drell--Yan pair production 
in nucleon--nucleon collisions \cite{Baldit:1994jk,Hawker:1998ty} 
have unambiguously shown that the light antiquark distributions 
in the proton are not flavor--symmetric, 
$\overline{d}(x) - \overline{u}(x) > 0$. The asymmetry
is clearly of non-perturbative origin and exhibits only weak
scale dependence, and therefore represents an interesting quasi--observable 
which directly probes the QCD quark structure of the nucleon at 
low resolution scales. The existence of a large flavor asymmetry
had been predicted \cite{Thomas:1983fh} on the basis of the
contribution of the nucleon's pion cloud to deep--inelastic
processes \cite{Sullivan:1971kd}. The ``bare'' nucleon couples
to configurations containing a virtual pion, and
transitions $p \rightarrow n \pi^+$ are more likely 
than $p \rightarrow \Delta^{++} \pi^-$, resulting in an excess of $\pi^+$
over $\pi^-$ in the ``dressed'' proton.
This basic idea was implemented in a variety of dynamical 
models, which incorporate finite--size effects
by hadronic form factors associated with the $\pi N N$ and 
$\pi N \Delta$ vertices; see Refs.~\cite{Kumano:1997cy}
for a review of the extensive literature. It was noted long 
ago \cite{FMS} that 
in order to reproduce the fast decrease of the observed
asymmetry with $x$ one needs $\pi NN$ form factors much softer than 
those commonly used in meson exchange parametrizations of the $NN$ 
interaction; however, even with such soft form 
factors pion virtualities (four--momenta squared)
generally extend up to values $\gg M_\pi^2$ \cite{Koepf:1995yh}. 
This raises the question to what extent such models actually describe 
long--distance effects associated with soft pion exchange 
(momenta $\sim M_\pi$), and what part of their predictions
is simply a parametrization of short--distance dynamics which 
should more naturally be associated with non-hadronic degrees of
freedom. More generally, one faces the question how to formulate the
concept of the ``pion cloud'' in the nucleon's partonic structure
in a manner consistent with chiral dynamics in QCD.

A framework which allows one to address these questions in a 
systematic fashion is the transverse coordinate (impact parameter)
representation, in which the distribution of partons is studied 
as a function of the longitudinal momentum fraction, $x$, and 
the transverse distance, $b$, of the parton
from the transverse center of momentum of the nucleon \cite{Burkardt:2002hr}. 
In this representation, chiral dynamics can be associated with 
a distinct component of the partonic structure, located at 
$x \lesssim M_\pi / M_N$ and $b \sim 1/M_\pi$ \cite{Strikman:2003gz}. 
In the gluon (more generally, in any flavor--singlet) 
distribution this large--distance component is sizable and 
contributes to the increase of the nucleon's average transverse 
size, $\langle b^2 \rangle$, with decreasing $x$ \cite{Strikman:2003gz}.
Here we study the large--distance component in the 
antiquark flavor asymmetry $\overline{d}(x) - \overline{u}(x)$
(\textit{i.e.}, the non-singlet sector). Using general
arguments, we first identify the region where parton distributions 
are governed by chiral dynamics. We then quantify what fraction 
of the asymmetry obtained in the phenomenological pion cloud model 
actually arises from this universal large--distance region. Finally,
we outline how the latter may be combined with a short--distance
``core'' in a two--component description of the nucleon's partonic 
structure. (A more detailed account of our studies will be given in a 
forthcoming article.)
\section{Chiral dynamics and partonic structure}
\label{sec:chiral}
The region where parton densities are governed by chiral dynamics
can be established from general considerations. The central object
is the pion longitudinal momentum and transverse coordinate distribution
in a fast--moving nucleon, $f_\pi (y, b)$ ($y$ is the pion momentum 
fraction); the precise meaning of this concept and its region
of applicability will be explained in the following.

\begin{figure}
\begin{tabular}{lcl}
\includegraphics[width=.3\textwidth]{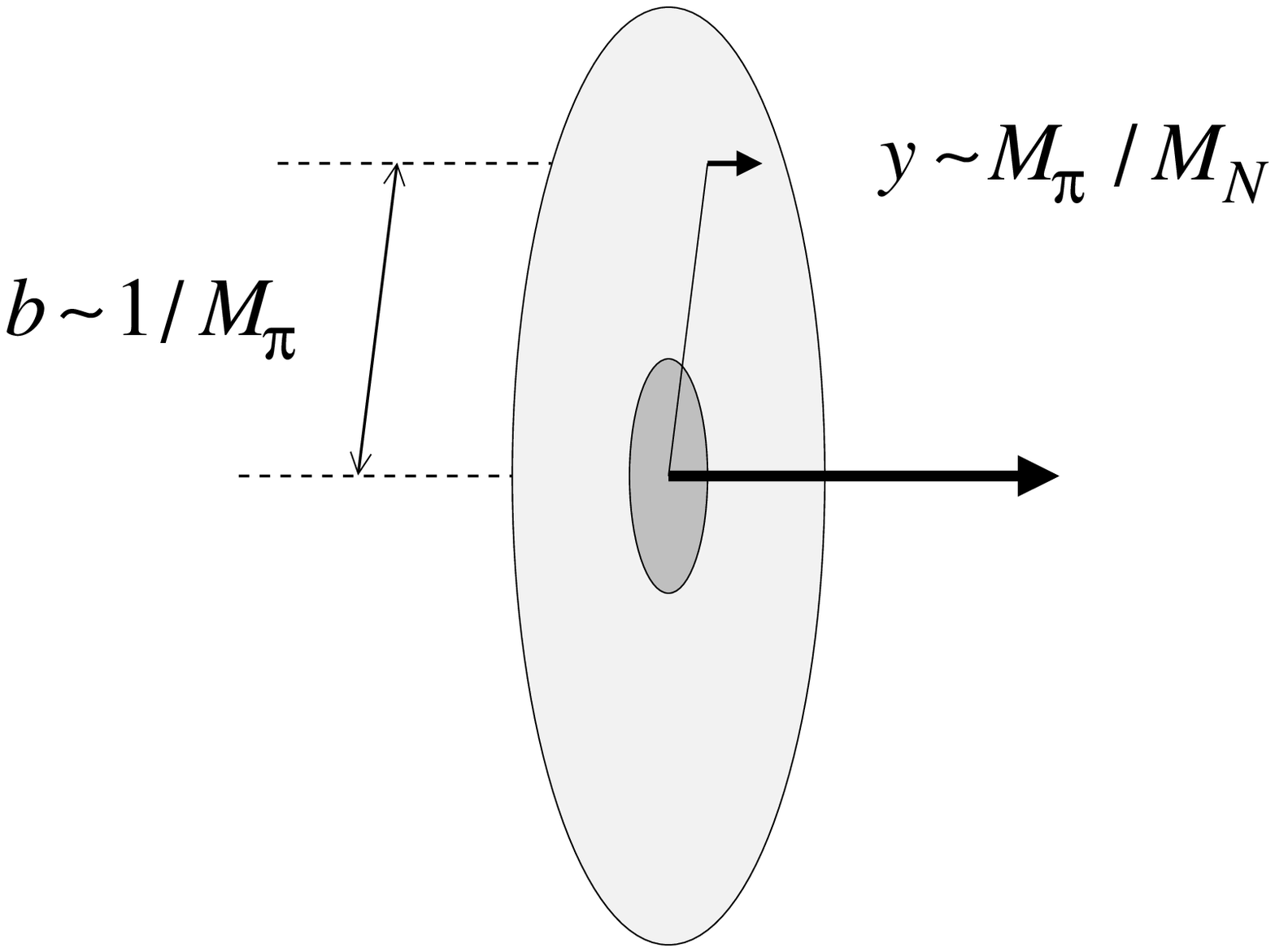}
& \hspace{.15\textwidth} &
\includegraphics[width=.32\textwidth]{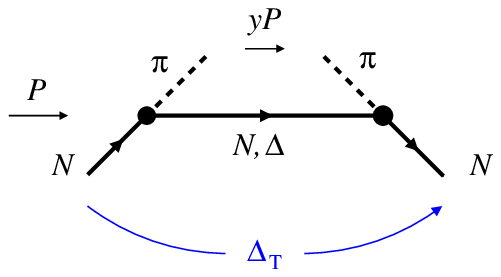}
\\[-2ex]
{\small (a)} & & {\small (b)}
\end{tabular}
\caption[]{(a) Parametric region where the pion distribution in the nucleon
is governed by chiral dynamics (in longitudinal momentum fraction $y$,
and transverse distance $b$). (b) The pion GPD in the 
nucleon. The distributions $f_{\pi B} (y, b) (B = N, \Delta)$ 
are obtained as the 2--dimensional Fourier transform 
$\bm{\Delta}_T \rightarrow \bm{b} \; (b \equiv |\bm{b}|)$.} 
\label{fig:chiral}
\end{figure}

Chiral dynamics generally governs long--distance contributions to 
nucleon observables, resulting from exchange of ``soft'' pions in the 
nucleon rest frame; in the time--ordered formulation of relativistic
dynamics these are pions with energies $E_\pi \sim M_\pi$ and momenta 
$|\bm{k}_\pi| \sim M_\pi$. Chiral symmetry provides that such pions 
couple weakly to the nucleon and to each other, so that their effects 
can be computed perturbatively. (The distance scale $1/M_\pi$ is assumed 
to be much larger than all other hadronic length scales in question.)
Boosting these weakly interacting $\pi N$ configurations 
to a frame in which the nucleon is moving with large velocity,
we find that they correspond to longitudinal pion momentum fractions
of the order\footnote{In covariant perturbation theory soft pions
have virtualities $-k_\pi^2 \sim M_\pi^2$, and Eq.~(\ref{y_chiral}) 
results from the condition that the minimum pion virtuality
required by the longitudinal momentum fraction, $y$, be of the 
order $M_\pi^2$.} 
\begin{equation}
y \;\; \sim \;\; M_\pi / M_N .
\label{y_chiral}
\end{equation}
At the same time, the soft pions' transverse momenta, which are not affected 
by the boost, correspond to a motion over transverse distances of the order 
\begin{equation}
b \;\; \sim \;\; 1/M_\pi .
\label{b_chiral}
\end{equation}
Together, Eqs.~(\ref{y_chiral}) and (\ref{b_chiral}) determine the 
parametric region where the pion momentum and coordinate distribution is 
governed by model--independent chiral dynamics, and the soft pion
can be regarded as a parton in the nucleon's wave function in the
usual sense (see Fig.~\ref{fig:chiral}a).
%
%

The condition Eq.~(\ref{y_chiral}) implies that the pion momentum
fraction in the nucleon is parametrically small, $y \ll 1$. 
As a consequence, one can generally neglect the recoil of the spectator 
system and identify the distance $b$ with the separation of the pion from 
the transverse center of momentum of the spectator system, 
$r = b/(1 - y)$. This makes for an important simplification in 
the spatial interpretation of chiral contributions to the parton densities.

In its region of applicability defined by Eqs.~(\ref{y_chiral}) 
and (\ref{b_chiral}), the $b$--dependent pion ``parton'' distribution 
can be defined unambiguously as the transverse Fourier transform 
of the pion GPD in the nucleon (see Fig.~\ref{fig:chiral}b).
The latter can be evaluated using either time--ordered or covariant 
perturbation theory, and implies summation over all relevant baryonic 
intermediate states. An important point is that, because
the pion wavelength is large compared to the typical
nucleon/baryon radius, only the lowest--mass excitations can 
effectively contribute to the GPD in the region of Eqs.~(\ref{y_chiral}) 
and (\ref{b_chiral}). We therefore retain only the $N$ and $\Delta$
intermediate states in the sum. The inclusion of the $\Delta$, whose
mass splitting with the nucleon introduces a non-chiral scale which is
numerically comparable to the pion mass, represents a slight departure 
from strict chiral dynamics but is justified by the numerical importance
of this contribution (\textit{cf.}\ the discussion of the
$N_c \rightarrow \infty$ limit in QCD below).
Calculation of the large--$b$ asymptotics 
of the pion distribution from the graph of Fig.~\ref{fig:chiral}b 
shows exactly the behavior described by 
Eqs.~(\ref{y_chiral}) and (\ref{b_chiral}) \cite{Strikman:2003gz}. 
For the $N$ intermediate state
\begin{equation}
f_{\pi N} (y, b) \;\; \propto \;\; e^{-\kappa_{\pi N}(y) b}/b, 
\hspace{3em} \;\;\; \textrm{with} \;\;\;
\kappa_{\pi N}(y) \; \sim \; M_\pi 
\;\;\; \textrm{for} \;\;\; y \; \sim \; M_\pi / M_N,
\end{equation}
\textit{i.e.}, for parametrically small $y$ the transverse distribution 
has a ``Yukawa tail'' with a $y$--dependent range of the order 
$\sim 1/M_\pi$ (the limiting value of $\kappa_{\pi N}$ for 
$y \rightarrow 0$ is $2 M_\pi$). 
For values $y \sim 1$ one finds an exponential decay with a range of the
order $\sim 1/M_N$, which is not a chiral contribution, in agreement
with the restriction Eq.~(\ref{y_chiral}).

The chiral contribution to the nucleon's parton densities is then
obtained as the convolution of the pion momentum distribution
in the nucleon with the relevant parton distribution in the pion.
For the antiquark flavor asymmetry it takes the form
(we suppress the QCD scale dependence)
\begin{equation}
\left[ \, \overline{d} - \overline{u} \right] (x, b)_{\rm chiral} 
\;\; = \;\; 
\int_x^1 \frac{dy}{y} \; 
\left[ \frac{2}{3} f_{\pi N} (y, b) - \frac{1}{3} f_{\pi \Delta} (y, b)
\right]
\; v_\pi (z), 
\hspace{3em} z \; \equiv \; x/y .
\label{conv}
\end{equation}
Here $f_{\pi N}$ and $f_{\pi \Delta}$ are the isoscalar pion 
distributions with $N$ and $\Delta$ intermediate states in the
conventions of Refs.~\cite{Koepf:1995yh,Strikman:2003gz}; 
the isovector nature of the asymmetry is encoded in the numerical 
prefactors. Furthermore, $v_\pi (z)$ denotes the ``valence'' 
quark/antiquark distribution in the pion,
\begin{equation}
v_\pi (z) \;\; = \;\; 
\left[ u_{\pi+} - \overline{u}_{\pi+} \right](z)
\;\; = \;\; 
\left[ \, \overline{d}_{\pi+} - d_{\pi+} \right](z),
\hspace{3em}
\int_0^1 dz \, v_\pi (z) \;\; = \;\; 1 .
\label{v_pi}
\end{equation}
Equation~(\ref{conv}) applies to quark momentum fractions of the order 
$x \lesssim M_\pi / M_N$, and transverse distances $b \sim 1/M_\pi$.
In arriving at Eq.~(\ref{conv}) 
we have assumed that the ``decay'' of the pion into quarks and antiquarks 
happens locally on the transverse distance scale of the chiral 
$b$--distribution, $b \sim 1/M_\pi$ (see Fig.~\ref{fig:chiral}a). 
This is justified if $x$ is not too small, as in this case the
antiquark momentum fraction $z$ in the pion does not reach
small values ($x < z < 1$ in the convolution integral) 
and one can neglect chiral effects which cause the transverse size of 
the pion itself to grow at small $z$; such effects were recently 
studied in Refs.~\cite{Kivel:2008ry} within a novel resummation
approach. 
\section{Impact parameter analysis of the pion cloud model}
We now turn to the phenomenological pion cloud model of the asymmetry
and investigate what part of its predictions actually correspond
to the long--distance region governed by universal chiral dynamics.
There are several variants of this model, employing different
types of form factors to regularize the transverse momentum
integral in the pion distribution (\textit{cf.}\ Fig.~\ref{fig:chiral}b).
We consider here the scheme in which the spectator baryon is
on mass--shell and the pion virtualities are restricted by 
form factors; the relation to other schemes (invariant mass cutoff
in the time--ordered approach) will be discussed elsewhere.

%
%
\begin{figure}
\begin{tabular}{ll}
\includegraphics[width=.48\textwidth]{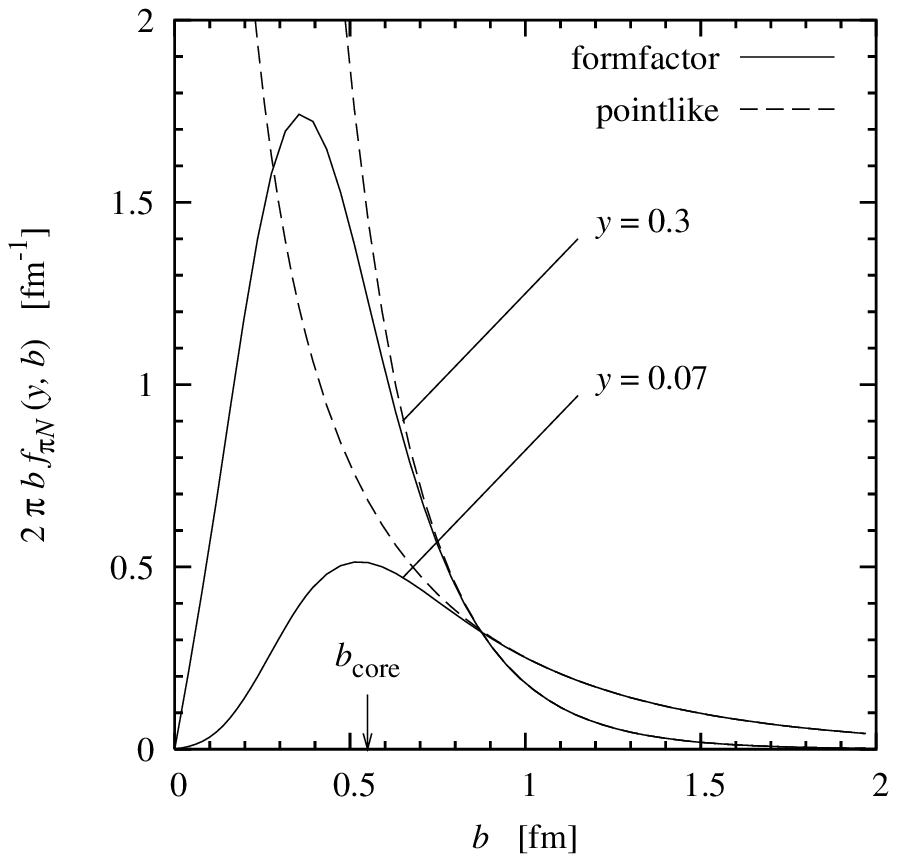} \hspace{-.6em}
& \hspace{-.6em}
\includegraphics[width=.48\textwidth]{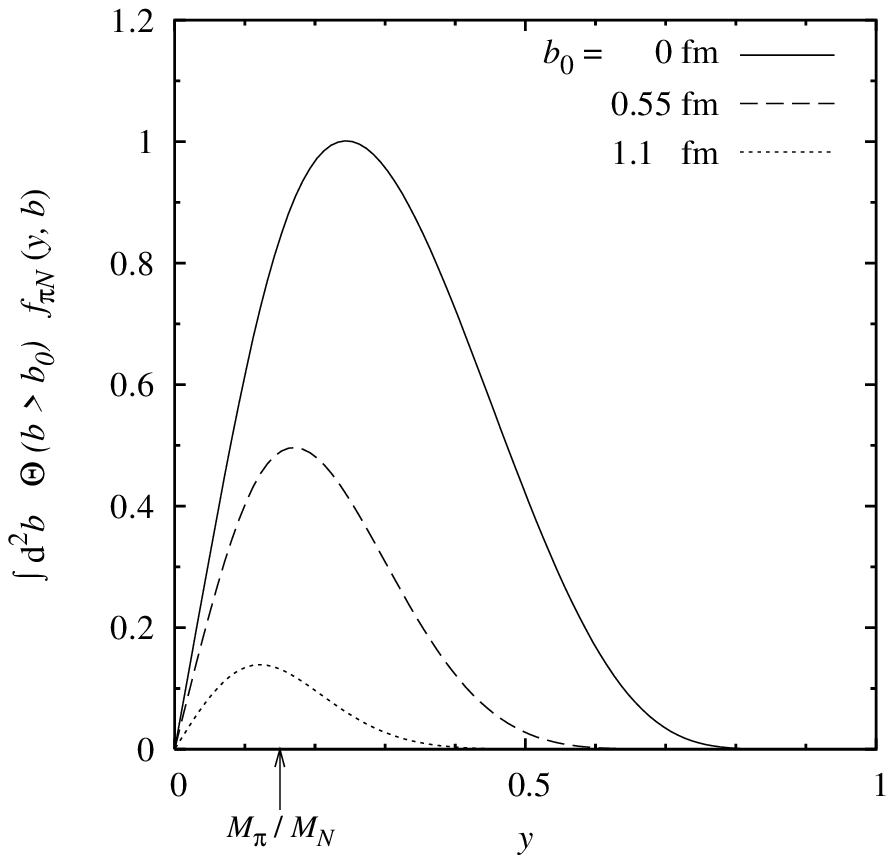}
\\[-2ex]
{\small (a)} & {\small (b)}
\end{tabular}
\caption[]{(a) The transverse spatial distribution of pions in 
the nucleon, $f_{\pi N} (y, b)$, as a function of $b$, for values
$y = 0.07$ and 0.3. Shown is the radial distribution $2\pi b \, f_{\pi N}$,
whose integral (area) gives the pion momentum 
distribution. \textit{Solid lines:} Pion cloud model with virtuality
cutoff (exponential form factor, $\Lambda_{\pi N} = 1.0 \, \textrm{GeV}$)
\cite{Koepf:1995yh}. \textit{Dashed line:} Distribution for pointlike
particles, regulated by subtraction at $\bm{\Delta}_T^2 = 0$
(the integral over $b$ does not exist in this case). 
(b) Momentum distribution of pions in the nucleon.
\textit{Solid line:} Full distribution $f_{\pi N}(y)$ obtained in pion 
could model (exponential form factor, $\Lambda_{\pi N} = 
1.0 \, \textrm{GeV}$). \textit{Dashed line:} Contribution from 
$b > b_{\rm core} = 0.55 \, \textrm{fm}$. 
Dotted line: Contribution from 
$b > 2 b_{\rm core} = 1.1 \, \textrm{fm}$.} 
\label{fig:fb}
\end{figure}
The pion momentum distributions $f_{\pi N, \pi\Delta}(y, b)$
in this model are calculated by evaluating the pion GPD from the
loop integral Fig.~\ref{fig:chiral}b with form factors, and performing the
transformation to the impact parameter representation. 
Figure~\ref{fig:fb}a shows $f_{\pi N}(y, b)$ as a function of
$b$ for $y = 0.07$ and 0.3 (which is 1/2 and 2 times 
$M_\pi / M_N$, respectively). Also shown are the distributions 
obtained with pointlike particles (no form factors), 
in which the loop integral was regularized by subtraction 
at $\bm{\Delta}_T^2 = 0$; this subtraction of a 
$\bm{\Delta}_T^2$--independent term in the GPD corresponds 
to a modification of the impact parameter distribution
by a delta function term $\propto \delta^{(2)}(\bm{b})$, 
which is ``invisible'' at finite $b$ \cite{Strikman:2003gz}. 
One sees that for $b \gtrsim 0.5 \, \textrm{fm}$ the results of the
two calculations coincide, showing that in this region the pion
distribution is not sensitive to finite size effects. It is 
interesting that the $b$ value where the universal behavior
sets in is numerically close to the transverse size of the 
``quark core'' inferred from the nucleon axial form factor, 
$b_{\rm core} = \left[ \frac{2}{3} \langle r^2 \rangle_{\rm axial} 
\right]^{1/2} \approx 0.55 \, \textrm{fm}$ \cite{Strikman:2003gz}.
This indicates that a two--component description of the partonic 
structure in transverse space, as advocated in Ref.~\cite{Strikman:2003gz}, 
is natural from the numerical point of view. Finally, we note
that for large $b$ both distributions in Fig.~\ref{fig:fb}a
exhibit the universal asymptotic behavior derived 
earlier \cite{Strikman:2003gz}.

We can now quantify which transverse distances contribute to the
pion momentum distribution in the pion cloud model with form factors.
Figure~\ref{fig:fb}b shows the momentum distribution of pions
obtained by integrating $f_{\pi N} (y, b)$ from a lower cutoff,
$b_0$, to infinity,
\begin{equation}
\int d^2 b \; \Theta (b > b_0) \; f_{\pi B} (y, b) 
\hspace{3em} (B = N, \Delta).
\label{fy_bint}
\end{equation}
Restricting the $b$ integration to values $b > 
b_{\rm core} = 0.55 \, \textrm{fm}$ strongly suppresses 
large pion momentum fractions and shifts the distribution
toward smaller $y$, in agreement with the general considerations
described in Sec.~\ref{sec:chiral}. Overall, we see that less than
half of the pions in the phenomenological pion cloud model arise from 
the ``safe'' long--distance region.
\section{Toward a two--component description}
\begin{figure}
\parbox[c]{.56\textwidth}{
\includegraphics[width=.56\textwidth]{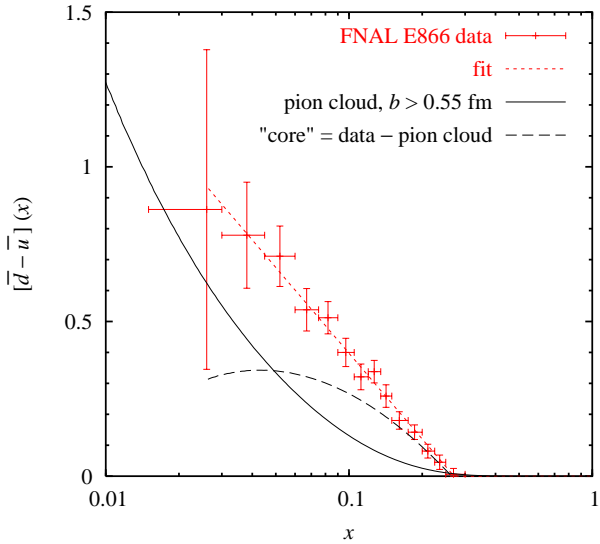}} 
\hspace{.02\textwidth}
\parbox[c]{.37\textwidth}{
\caption[]{Spatial decomposition of the flavor asymmetry
$[\overline{d} - \overline{u}](x)$. The data are from the FNAL E866 
Drell--Yan experiment \cite{Hawker:1998ty}. \textit{Dotted line:}
Empirical fit to the data. \textit{Solid line:} Large--distance 
component calculated in the pion cloud model ($b > b_{\rm core} \
= 0.55 \, \textrm{fm}$). \textit{Dashed line:} ``Core'' component, 
obtained as the difference of the calculated large--distance component
and the fit to the data.
(The cutoff parameters in the pion cloud model were chosen 
as $\Lambda_{\pi N} = 1.0 \, \textrm{GeV}, 
\Lambda_{\pi\Delta} = 0.8 \, \textrm{GeV}$ \cite{Koepf:1995yh},
and the valence quark density in the pion was taken from 
Ref.~\cite{Gluck:1999xe}.)}
\label{fig:conv}}
\end{figure}

The above results allow us to make a first attempt at a spatial
decomposition of the antiquark flavor asymmetry in the proton.
To this end, we calculate the convolution integral Eq.~(\ref{conv})
with the $b$--integrated pion distribution Eq.~(\ref{fy_bint}),
where the lower limit, $b_0$, is taken sufficiently large to 
exclude the model--dependent short--distance region
(see Fig.~\ref{fig:fb}a). Figure~\ref{fig:conv}
(solid line) shows the result obtained with $b_0$ taken as
the phenomenological ``core'' radius, $b_{\rm core} = 0.55 \, \textrm{fm}$.
Also shown in the figure are the data from the FNAL E866 experiment
\cite{Hawker:1998ty} and an empirical fit to the data (dotted line). 
Taking the difference between the fit to the data and the calculated 
long--distance contribution, we can infer the asymmetry in the ``core'' 
in a quasi model--independent way (dashed line). It is interesting 
that the ``core'' component of $\overline{d}(x) - \overline{u}(x)$ defined
in this way seems to tend to zero for $x \rightarrow 0$; this is 
the behavior one would expect of antiquark densities obtained from 
the overlap of soft light--cone wave functions with non--minimal
$\overline{q} q$ components (however, the quality of the
data is rather poor at small $x$, and we cannot reliably 
extrapolate to smaller $x$ at present). Nevertheless, our results 
show that a generic two--component model, with a quark core of transverse 
size $b_{\rm core} \approx 0.55\, \textrm{fm}$ and a universal chiral
long--distance contribution, should be able to describe the 
$x$--dependence of the measured asymmetry rather naturally.

We note that without the restriction to $b > b_{\rm core}$ 
the pion cloud model with virtuality cutoff, with the above parameters, 
would give an asymmetry which overshoots the data at small $x$ and 
is non-zero also for $x > 0.3$ (this could partly be remedied 
by tuning the cutoff parameters). The two--component description 
in transverse coordinate space proposed here solves this
problem in a natural --- and theoretically more satisfying --- way.
%
%
\section{Summary and discussion}
The transverse coordinate representation based on GPDs is
a most useful framework for studying the role of chiral dynamics in
the nucleon's partonic structure. Parametrically, the chiral
contributions are located at momentum fractions $x \lesssim M_\pi / M_N$
and transverse distances $b \sim 1/M_\pi$. We have shown that the
results of the phenomenological pion cloud model become independent 
of the $\pi NN$ form factors at transverse distances 
$b \gtrsim 0.5 \, \textrm{fm}$ and represent contributions
governed by universal long--distance dynamics. The lower limit
in $b$ approximately coincides with the ``quark core'' radius
$b_{\rm core} = 0.55 \, \textrm{fm}$, inferred previously from
other phenomenological considerations, suggesting a natural 
``two--component'' description of the partonic structure
in transverse coordinate space \cite{Strikman:2003gz}. 
Interestingly, the present data on $\overline{d}(x) - \overline{u}(x)$
are well described by complementing the universal large--distance 
contribution with a ``valence--like'' core distribution, as would be
obtained from soft light--cone wave functions with $\overline{q} q$
components. Our findings provide a starting point for more
detailed modeling of the nucleon's partonic structure along 
these lines. 

The pion cloud contribution to the flavor asymmetry 
$\overline{d}(x) - \overline{u}(x)$ 
involves strong cancellations between $\pi N$ and $\pi \Delta$ intermediate 
states. A systematic way to deal with this problem is provided by the 
$1/N_c$ expansion of QCD. In particular, the degeneracy of $N$ and $\Delta$ 
in the $N_c \rightarrow \infty$ limit ensures the proper $1/N_c$ scaling 
of the pion cloud contribution to the flavor asymmetry
\cite{Strikman:2003gz}, showing that the latter is a legitimate part of 
the nucleon's partonic structure in large--$N_c$ QCD. The connection
between the pion cloud contribution to $\overline{d}(x) - \overline{u}(x)$ 
and the dynamical picture of the nucleon as a chiral soliton in the 
large--$N_c$ limit remains an interesting problem 
for further study \cite{Strikman:2003gz}.

Notice: Authored by Jefferson Science Associates, LLC under U.S.\ DOE
Contract No.~DE-AC05-06OR23177. The U.S.\ Government retains a
non-exclusive, paid-up, irrevocable, world-wide license to publish or
reproduce this manuscript for U.S.\ Government purposes.

\end{document}